\documentclass[10pt,conference]{IEEEtran}
\IEEEoverridecommandlockouts
\usepackage{graphicx}
\usepackage{wrapfig}
\usepackage{amsmath}
\usepackage{algpseudocode}
\usepackage{algorithm}
\usepackage{pgfplots}
\pgfplotsset{compat=1.10}
\pgfplotsset{width=8.3cm,compat=1.9}
\usepackage{multirow}
\usepackage{caption}
\usepackage{float}
\usepackage{enumerate}
\usepackage[left=1.62cm,right=1.62cm,top=1.9cm,bottom=2.4cm]{geometry}

\graphicspath{ {./images/} }

\pgfplotsset{compat=1.18}
\begin{document}

\title{Secure and Efficient Entanglement Distribution Protocol for Near-Term Quantum Internet \\\

\vspace{-10mm}

\author{Nicholas Skjellum$^\S$, Mohamed Shaban$^{\S,\dagger}$ and Muhammad Ismail$^\S$\\

{$^\S$Department of Computer Science, Tennessee Technological University, Cookeville, TN, USA}\\
{$^\dagger$Department of Mathematics, Faculty of Education, Alexandria University, Egypt} \\

Emails: \{nskjellum42, mmibrahims42, mismail\}@tntech.edu

\thanks{This work was supported by NSF Award \#2210251.}
}
}

\maketitle

\begin{abstract}
Quantum information technology has the potential to revolutionize computing, communications, and security. To fully realize its potential, quantum processors with millions of qubits are needed, which is still far from being accomplished. Thus, it is important to establish quantum networks to enable distributed quantum computing to leverage existing and near-term quantum processors into more powerful resources. This paper introduces a protocol to distribute entanglements among quantum devices within classical-quantum networks with limited quantum links, enabling more efficient quantum teleportation in near-term hybrid networks. The proposed protocol uses entanglement swapping to distribute entanglements efficiently in a butterfly network, then classical network coding is applied to enable quantum teleportation while overcoming network bottlenecks and minimizing qubit requirements for individual nodes. Experimental results show that the proposed protocol requires quantum resources that scale linearly with network size, with individual nodes only requiring a fixed number of qubits. For small network sizes of up to three transceiver pairs, the proposed protocol outperforms the benchmark by using $17\%$ fewer qubit resources, achieving $8.8\%$ higher accuracy, and with a $35\%$ faster simulation time. The percentage improvement increases significantly for large network sizes. We also propose a protocol for securing entanglement distribution against malicious entanglements using quantum state encoding through rotation. Our analysis shows that this method requires no communication overhead and reduces the chance of a malicious node retrieving a quantum state to $7.2\%$. The achieved results point toward a protocol that enables a highly scalable, efficient, and secure near-term quantum Internet.
\end{abstract}

\IEEEpeerreviewmaketitle

\section{Introduction}

 \IEEEPARstart{Q}{uantum} information technology represents a key enabler to new applications within computing, communications, sensing, intelligence, and security. However, the world's most powerful quantum processor, the IBM Osprey, has only $433$ qubits, while the second most powerful processor from IBM, the IBM Eagle, has only $127$ qubits. To effectively realize the applications of quantum computing, a processor that holds millions of qubits is necessary, which is far from being accomplished in the near term. Instead, distributed quantum computing using existing processors can achieve this potential in the near term. 
 Hence, a quantum Internet capable of connecting multiple quantum processors across large distances is required to enable distributed quantum computing. To facilitate a near-term implementation of distributed quantum networks, we expect that the early versions will consist of hybrid classical-quantum networks, where a central node or nodes will be connected to all other quantum devices through optical fiber network links (compatible with quantum and classical communications), while the remaining devices will be connected with each other through the classical Internet. To enable this network structure, quantum teleportation \cite{PhysRevLett.70.1895}, a technique that utilizes classical bits to transmit quantum states, can be used to allow these quantum devices to exchange quantum states without having a direct quantum-compatible link. However, to utilize quantum teleportation, entanglements are required. Entanglements must be generated locally within a single device (central node or nodes), with the entangled pair halves then distributed to all devices participating in the quantum teleportation. As a result, the means to create, distribute, and exploit these entanglements \cite{concurrent-eqc} in an efficient manner is integral to the development of a quantum Internet. 
 

To maximize the benefits of teleportation on a hybrid network, entanglement distribution must efficiently utilize the limited number of quantum links. When direct quantum links between nodes are unavailable, nodes will rely on alternate paths to distribute entanglements, which often creates bottlenecks shared by multiple communicating parties, hence, limiting the teleportation throughput. While there exists research that optimizes quantum-classical network structures \cite{hybrid-qc-nw-op}, we seek to provide a solution that efficiently utilizes the existing network structure to better distribute entanglements and improve the teleportation throughput. As the best way to demonstrate bottlenecks is through a butterfly network structure, our proposed solution is investigated on a butterfly network.

Additionally, as entanglements are distributed through the network, individual devices may not be trusted to assist in the entanglement distribution process. Malicious quantum devices can perform malicious entanglement by secretly entangling themselves with the communicating nodes, resulting in a man-in-the-middle attack. Whenever these nodes communicate by teleportation, the malicious device will be able to eavesdrop on the teleported quantum state with $100\%$ probability as long as it is able to discover the $2$-bit classical string or otherwise by a $25\%$ probability with a random guess. Existing methods to secure teleportation, such as controlled teleportation, do not inherently secure the quantum states being transmitted and do not provide a solution for this issue. To combat this and other types of eavesdropping, we seek to provide protection to teleportation without increasing the classical communication overhead. Therefore, this paper aims to develop an efficient and secure entanglement distribution protocol that overcomes bottleneck links to enable an optimal teleportation rate between indirectly connected quantum devices.

\subsection{Related Works}

In \cite{qnw-remote} and \cite{qnw-r-state}, quantum network coding is used to facilitate the preparation of multi-qubit and qudit states, respectively, between indirectly connected nodes. In \cite{anti-noise-q}, quantum network coding is utilized to reduce noise within communicated messages. The work in \cite{demonstration} proposes a method for distributing measurement-based entanglements by creating a fully entangled butterfly network. However, all these works require the existence of a full optical fiber-connected network in order to perform quantum network coding, making them incompatible with the near-term hybrid quantum Internet. The work in \cite{q-linear-nw} suggests specific network structures that provide greater efficiency but do not inherently address throughput limits on network bottlenecks. For securing quantum communication, literature such as \cite{secure-qnw} forgoes classical communications and utilizes only quantum information, which makes it inapplicable for near-term hybrid classical-quantum networks. The work in \cite{single-shot-qnw} approaches quantum security by only allowing unidirectional communication, which does not require the distribution of entanglements, but as other research, requires a fully quantum network and is not compatible with our research goals. 

Much of the existing literature does not address entanglement distribution on classical-quantum networks, nor do their solutions scale efficiently even when distributing quantum states. Alternative configurations fail to provide a general solution that can be employed in any network configuration. Furthermore, the security solutions presented in the literature are only applicable to quantum-only communications and are prohibitive when used to protect the entanglement distribution or otherwise make it difficult to perform. 

One work, \cite{herbert_ghz}, proposes a protocol that seeks to overcome bottleneck links while supporting entanglement distribution between indirect nodes, and is implemented in a classical-quantum network. We utilized this protocol as a benchmark to evaluate the efficiency of our proposed solution. 

\subsection{Contributions}

To support a near-term quantum Internet on a classical-quantum network we contribute the following: 
\begin{itemize}

    \item We propose IEDTC (Indirect Entanglement Distribution with Teleportation Coding), a protocol that utilizes entanglement swapping between nodes to efficiently distribute entanglements over bottlenecks for teleportation. To deliver the teleported quantum states, classical messages are then sent to each receiver node through classical network coding, hence, teleporting quantum states and avoiding throughput limitations due to bottlenecks. 

    \item We propose QSRE (Quantum State Rotation Encoding) to protect against malicious entanglements by encoding quantum states using angle rotation. Using a pre-shared classical private key, transceiver pairs teleport quantum states securely and deny eavesdroppers the contents of a teleported state. The encoded states do not expose the contents of the classical key, allowing key reuse by both parties, without classical communication overhead.

    \item We implement IEDTC, QSRE, and the benchmark \cite{herbert_ghz} in a quantum network simulator, QuNetSim \cite{diadamo2020qunetsim}, and evaluate the results of the simulation.     
\end{itemize}

The rest of this paper is as follows. Section \ref{sec:sys_model} presents the network model. Section \ref{sec:proposed_protocol} proposes the IEDTC protocol. Section \ref{sec:security_pros} proposes the QSRE protocol. Section \ref{sec:results} provides the simulation results. Section \ref{sec:conclusions} concludes the paper.
\section{System Model}
\label{sec:sys_model}
\begin{figure}
    \centering
    \includegraphics[width=0.25\textwidth]{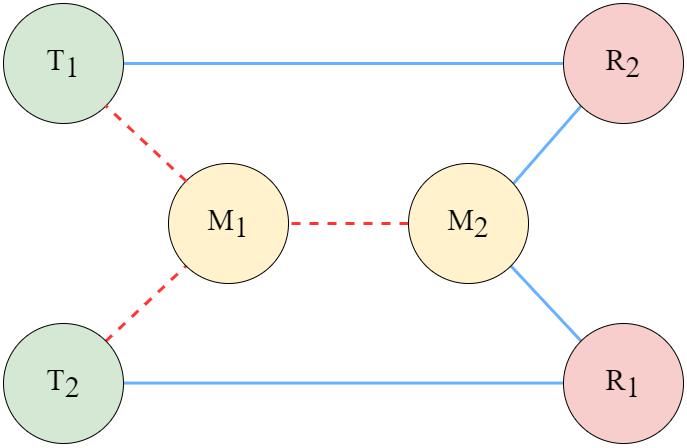}
    \caption{Butterfly classical-quantum network. Blue connections represent optical fiber links. Red connections represent classical links. The link between $M_1$ and $M_2$ is the bottleneck link.}
    \label{fig:iedtc-bnw}
\end{figure}

Consider a hybrid classical-quantum butterfly network as an example to demonstrate entanglement distribution with limited quantum links and bottlenecks. The butterfly network consists of a set of transmitter nodes, each denoted by $T_n$, and receiver nodes, each denoted by $R_n$, with each $T_n$ and $R_n$ referred to as a transceiver pair. In this network, neither node of any transceiver pair is directly connected with one another. We define the network entities below, with a visual depiction in Fig. \ref{fig:iedtc-bnw} for a size $2$ network (i.e., $2~T_n-R_n$ pairs).

\begin{itemize}
\item Each transmitter $T_n$ is connected to all receivers in the network, except $R_n$, through optical fiber links.
\item Each transmitter $T_n$ is connected to a central node $M_1$ via a classical link. Each transmitter $T_n$ is a source of entanglement. 
\item Each receiver $R_n$ is connected to a central node $M_2$ via an optical fiber link. $M_2$ is a source of entanglement.
\item The central nodes $M_1$ and $M_2$ are connected through a classical link, which represents the network bottleneck.
\end{itemize}

\begin{figure}[!t]
\centering
\includegraphics[width=0.40 \textwidth]{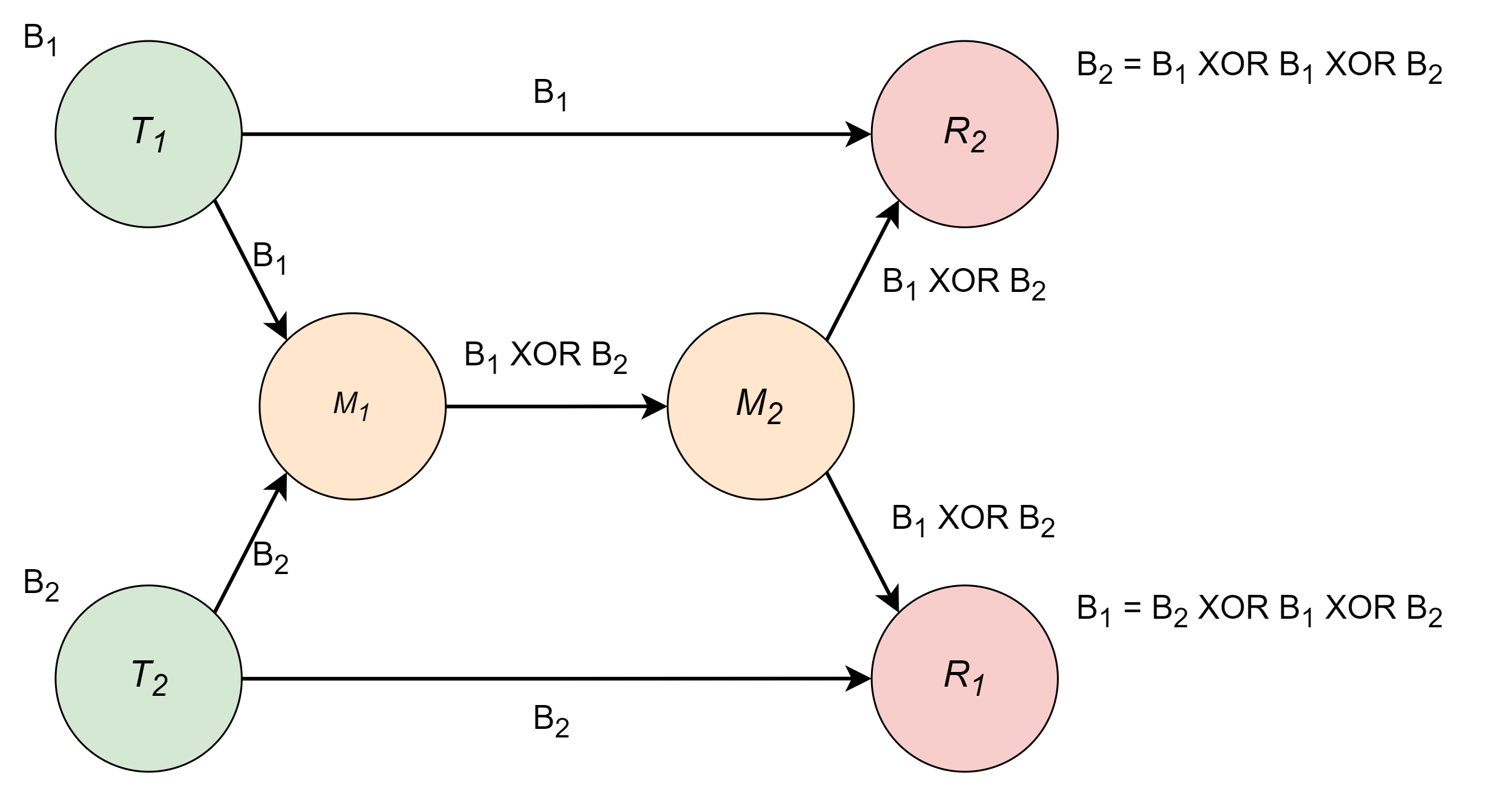}
\caption{Illustration of classical network coding to increase the throughput when there is a network bottleneck.}
\label{fig:xor_nw}
\end{figure}

For classical messages, classical network coding \cite{850663}, depicted in Fig. \ref{fig:xor_nw}, is employed to circumvent the network bottleneck by sending multiple classical messages $B_n$ simultaneously through the same middle link between $M_1$ and $M_2$. Each transmitter broadcasts its message to all connected receivers and $M_1$. Then, $M_1$ XORs all the messages into a single string and sends it to $M_2$. Next, $M_2$ broadcasts this string to each receiver. Finally, each receiver recovers the original message sent by its transmitter pair by XORing the string sent by $M_2$ with the messages received from the directly connected transmitters, hence, recovering $B_n$. 

When teleporting quantum states, two steps are required: (1) distributing entanglements between each transmitter-receiver pair and (2) exchanging classical teleportation messages between each transmitter-receiver pair to recover the teleported state. For the classical messages (step 2), classical network coding can be used to overcome the bottleneck between $M_1$ and $M_2$. However, classical network coding techniques cannot be applied when distributing the entangled quantum pairs through the bottleneck between $M_1$ and $M_2$ (step 1) due to the no-cloning theorem. Hence, we aim to develop a protocol to distribute the entangled states while overcoming the bottleneck.

\section{IEDTC: Proposed Indirect Entanglement Distribution with Teleportation Coding Protocol}
\label{sec:proposed_protocol}

This section introduces the proposed IEDTC protocol for entanglement distribution over network bottlenecks. IEDTC utilizes entanglement swapping and quantum teleportation to transmit quantum states across the network. Entangled pairs are generated in a central node and in each transmitter. Pair halves are then distributed to connected receivers. These receivers perform entanglement swapping to provide each transmitter with an entangled state that is shared by its desired receiver. With entanglements established between transceiver pairs, quantum states are sent using quantum teleportation, which transmits a classical message from the transmitter to its paired receiver. Here, classical network coding is used to efficiently transmit multiple teleportation messages and overcome the network bottleneck. Without loss of generality, the following example demonstrates the protocol on a size $2$ network, as depicted in Fig. \ref{fig:iedtc-bnw}, and the same principles are applied for larger sizes: 

\begin{enumerate}
    \item Transmitter $T_1$ creates an entangled pair $|\phi_{1}$⟩ $=|\phi_{11}$⟩ $|\phi_{12}$⟩ and sends half of it ($|\phi_{11}$⟩) to $R_2$. Similarly, transmitter $T_2$ creates an entangled pair $|\phi_{2}$⟩ $=|\phi_{21}$⟩ $|\phi_{22}$⟩ and sends half of it ($|\phi_{21}$⟩) to $R_1$.  
    
    \item Central node $M_2$ creates two entangled pairs, $|\psi_{1}$⟩ $= |\psi_{11}$⟩ $|\psi_{12}$⟩ and $|\psi_{2}$⟩ $= |\psi_{21}$⟩ $|\psi_{22}$⟩. Then, $M_2$ distributes the halves of $|\psi_{1}$⟩, sending $|\psi_{12}$⟩ to $R_1$ and $|\psi_{11}$⟩ to $R_2$. Similarly, it distributes the halves of $|\psi_{2}$⟩ and sends $|\psi_{21}$⟩ to $R_1$ and sends $|\psi_{22}$⟩ to $R_2$. 
    
    \item Node $R_1$ now holds $|\phi_{21}$⟩ (entangled with $T_2$), $|\psi_{12}$⟩ (entangled with $R_2$), and $|\psi_{21}$⟩ (entangled with $R_2$). Then, $R_1$ consumes $|\phi_{21}$⟩ to teleport $|\psi_{21}$⟩ to $T_2$, thus performing entanglement swapping. Now, $T_2$ and $R_2$ are entangled. In this step, $R_1$ is serving as an entanglement swapping node for $T_2$ and $R_2$.   

    \item Likewise, $R_2$ holds $|\phi_{11}$⟩ (entangled with $T_1$), $|\psi_{11}$⟩ (entangled with $R_1$), and $|\psi_{22}$⟩ (entangled with $R_1$). Then, $R_2$ consumes $|\phi_{11}$⟩ to teleport the quantum state $|\psi_{11}$⟩ to $T_1$, performing an entanglement swap as well. Now, $T_1$ and $R_1$ are entangled. In this step, $R_2$ is serving as an entanglement swapping node for $T_1$ and $R_1$.   
\end{enumerate}

\begin{figure}
    \centering
    \includegraphics[width=0.35\textwidth]{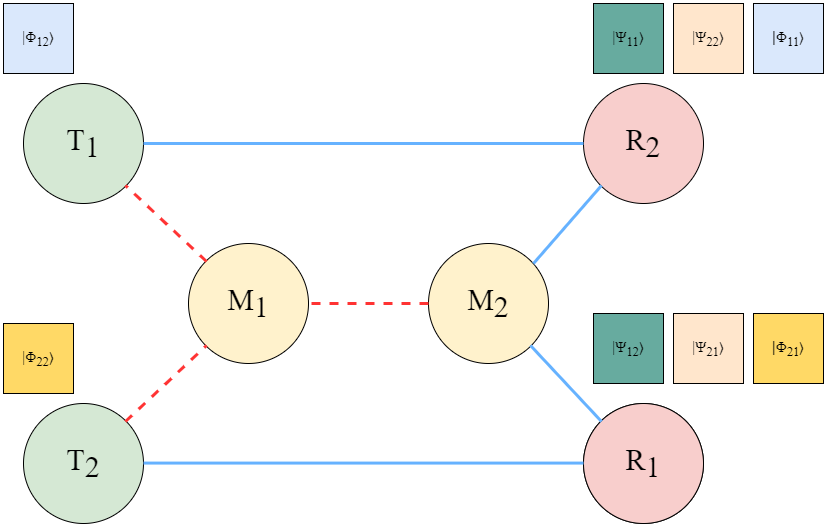}
    \caption{Entangled states prior to entanglement swapping.}
    \label{fig:relation}
\end{figure}

\begin{figure}
    \centering
    \includegraphics[width=0.30\textwidth]{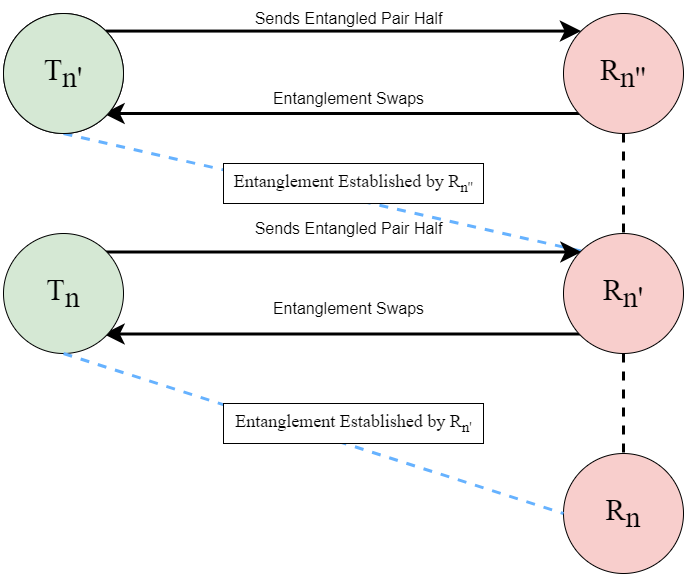}
    \caption{Relationships between receivers in a larger network.}
    \label{fig:n-rel}
\end{figure}

As shown in Fig. \ref{fig:relation}, a given receiver $R_{n'}$ will assist in entangling an adjacent receiver $R_n$, where $n \neq n'$, with its transmitter pair $T_n$. In a general setting, Fig. \ref{fig:n-rel} depicts the expanded relationship for a network of size $N$. Each receiver entangles an adjacent receiver with its paired transmitter, and this receiver in turn is entangled with its paired transmitter by another receiver. Each receiver distributes entanglements in parallel with all other receivers, thus, increasing the efficiency of the protocol. Also, the independence of each distributed entanglement allows central node $M_2$ to distribute entanglements on demand rather than as a batch, hence, reducing the resource requirements of this node.

With entanglements distributed, each transceiver pair ($T_n - R_n$) holds half of an entangled pair, which is then used to perform quantum teleportation of any state $|\eta_n$⟩. As part of teleportation (step 2 discussed in Section II), a $2$-bit classical string $B_n$ is created, and hence, classical network coding is used to transmit this teleportation message. Specifically, each transmitter $T_n$ performs a Bell measurement on its entangled half and the qubit containing $|\eta_n$⟩, generating $B_n$. Then, each transmitter broadcasts its teleportation message to all connected receivers and $M_1$. Next, $M_1$ will combine all teleportation messages it receives into a single classical message using a logical-XOR operation, then transmit this combined message to $M_2$, which $M_2$ broadcasts to all connected receivers. Then, each receiver subtracts each teleportation message $B_{n'}$ it received from each $T_{n'}$ from the combined message using an additional XOR operation. This recovers $B_n$, which $R_n$ uses as control values for quantum gates to retrieve $|\eta_n$⟩. Using our proposed entanglement distribution approach based on $M_2$ and classical network coding based on $M_1$, transceiver pairs can perform teleportation while overcoming the bottleneck link $M_1 - M_2$ and the limited number of quantum links.

\section{Proposed Security Protocol}
\label{sec:security_pros}

Inherently trusting nodes to assist with teleportation as in the IEDTC protocol exposes the network to eavesdropping through malicious entanglement \cite{9477172}. To eavesdrop in this manner, a node $R_n$, which is directly connected to transmitter $T_{n'}$, attempts to acquire a quantum state that would originally be teleported to $R_{n'}$. To do this, $R_n$ creates an entangled pair and transmits half of it to $T_{n'}$. Instead of assisting in creating a shared entanglement between $R_{n'}$ and $T_{n'}$, there is instead an entanglement between $R_n$ and $T_{n'}$. In this situation, the malicious node is easily capable of retrieving any quantum state teleported by $T_{n'}$. This jeopardizes (a) the privacy of the network, as no quantum state is safe from being eavesdropped, and (b) the availability of the network, as $R_{n'}$ is unable to retrieve its intended quantum state. 

To combat this threat, we introduce Quantum State Rotation Encoding (QSRE) to secure any teleported quantum state. A quantum state can be described as being on the surface of the Bloch sphere. We can encode the quantum state by rotating it without requiring additional classical bits when sending a teleportation message. If a malicious node were to eavesdrop on this encoded state, it would need to perform the inverse rotation performed on the original quantum state. Assuming that $R_n$ and $T_n$ are sharing a private key $S_n$, we outline the encoding process of QSRE as follows:


\begin{enumerate}
    \item A node $T_n$ wants to teleport a quantum state $|\eta_n$⟩. Before teleporting this state, the transmitter $T_n$ will read the $i$th $2 + D$ bits from a pre-shared private key $S_n$ that it shares with $R_n$, where $i$ is the message number in a round of communications, and $D$ is the number of bits used to reflect the degrees of rotation. 
    
    \item $T_n$ will use the first bit to rotate in either the X-axis or Y-axis, with a $0$ rotating the state around the X-axis, and $1$ the Y-axis. $T_n$ will then use the second bit to determine the direction, negative or positive, of this rotation. 
    \item The last $D$ bits will be used to determine the angle of rotation, using the following equation: $\theta$ \( \displaystyle = \frac{\pi}{-1^{b} \times (1+ d)} \), where $b$ is the value of the second bit used to determine the direction of the rotation, and $d$ is the decimal value of the $D$ bits. For example, if the bits read from the private key were $1011$, the quantum state would be rotated around the Y-axis, in the negative direction, with a decimal value of $d$ evaluating to 3. 
    
    \item Once rotated, the quantum state $|\eta_n$⟩ will become $|\eta_{n}'$⟩, which will then be teleported. To recover $|\eta_n$⟩, $R_n$ will retrieve $|\eta_{n}'$⟩, then read the same $i$th bits from its shared private key $S_n$ and perform the inverse of the rotation performed by $T_n$, recovering $|\eta_n$⟩.
\end{enumerate}

For a malicious receiver to recover $|\eta_n$⟩, it must guess the axis of rotation, the direction, and the magnitude of the rotation angle $\theta$. The ideal (lowest) success rate of a malicious receiver attempting to defeat this method is \( \displaystyle \frac{1}{2^{D+2}}\). Also, encoding a quantum state in this manner reveals no information from the private key to an eavesdropper, hence, allowing key reuse. 

By integrating IEDTC with QSRE, secure and efficient teleportation can be accomplished across hybrid classical-quantum links with bottlenecks and a limited number of quantum links. 

\section{Results and Discussion}
\label{sec:results} 

To assess the performance of IEDTC and QSRE in QuNetSim, the following metrics have been considered. For IEDTC and the benchmark protocol \cite{herbert_ghz} we examined qubit usage, the number of links required for the network of a given network size $N$, the simulation time, and the accuracy of the protocol when noise was introduced to the network. 


Qubit usage is calculated by the maximum number of qubits containing unique quantum states that are required to be maintained or utilized at a given time. A given node could possibly see a greater number of states than qubits required to function within the protocol. The number of required links is determined by the number of network connections across the entire network between nodes. 
Simulation time is the time required for the protocol to be completed within QuNetSim by our implementation, consisting of entanglement distribution and quantum teleportation phases with IEDTC and QSRE. Accuracy is the success rate of a protocol when quantum gates had an $X\%$ chance to introduce noise. 

\subsection{Setup}
QuNetSim was configured within an Ubuntu distribution for the Windows Linux Subsystem, using the backend EQSN \cite{eqsn} that was provided with QuNetSim. The environment was run with 16 cores with a speed per CPU of $2419.198$ MHZ. To check for errors in quantum states, EQSN was modified to provide state vectors of qubits within the network. To implement noise, the backend was modified so that quantum gates would introduce noise at a predetermined rate, with an increase in the rate introducing noise more frequently into the network whenever a quantum gate was utilized. Accuracy was tested over a range from $1\%$ to $10\%$ chance to introduce noise. Network size was determined by the number of transceiver pairs ($T_n - R_n$) within the network. The benchmark implementation of \cite{herbert_ghz} had a limited network size, with sizes larger than $3$ not possible due to a memory exception raised by the numpy library: "\textit{numpy.core.\textunderscore exceptions.\textunderscore ArrayMemoryError}," with QuNetSim requiring a 2-dimensional array of float$64$ integers of $512$ gigabytes to simulate a network size of $4$, with requirements increasing for larger network sizes. We did not possess a machine capable of providing this amount of memory for our simulations and were thus unable to test the benchmark for network sizes greater than $2$.

\subsection{IEDTC Results}
 Table \ref{tab:resouce-table} shows the resource costs of IEDTC and the benchmark. Link usage increases quadratically for both protocols, but the benchmark requires more as it requires additional links in its network compared to IEDTC. Of greater note is that IEDTC has a linearly increasing qubit usage as the network size increases, while the benchmark has a quadratically increasing qubit usage, with IEDTC requiring $17\%$ fewer qubits overall.

 \begin{table}
\begin{tabular}{ |c|c|c|c|  }
 \hline
 Protocol &
  \begin{tabular}[c]{@{}l@{}}Total Number \\ of Links\end{tabular} &
  \begin{tabular}[c]{@{}l@{}}Number of \\ Quantum Links\end{tabular} &
  \begin{tabular}[c]{@{}l@{}}Number of \\ Qubits\end{tabular}\\
 \hline
     \cite{herbert_ghz} & \(3N^2 + 5N + 2\) & \(2N^2 + 5N + 2\) & \(2N^2+3N + 1\) \\
    IEDTC & \(N^2 + N + 1\) & $N^2$ &\(7N\) \\
 \hline

\end{tabular}
    \caption{Resource usage comparison between IEDTC and benchmark \cite{herbert_ghz}.}
    \label{tab:resouce-table}

 \end{table}

To test accuracy, each protocol was run over $1000$ trials at each noise probability. A trial's success required all states to be correctly received, any incorrect states resulted in a failed trial. We determined correctness by comparing the original and transmitted quantum states from each trial. Successes and failures were averaged for each level of noise with a $95\%$ confidence interval. Fig. \ref{fig:accuracy} compares the accuracy of both protocols. 
 Experiments show that IEDTC and the benchmark \cite{herbert_ghz} had comparable accuracy at a network size $2$, but IEDTC performed with better accuracy at a network size $3$.  Two factors contributed to this. Firstly, the benchmark utilizes an extensive entanglement containing all qubits within the network, with any noise introduced propagated to all other quantum states. Secondly, the greater number of qubits within the benchmark at network size $3$ required more quantum operations to distribute entanglements, providing more chances for a gate operation to introduce noise. In IEDTC, any error introduced was contained to a specific entanglement. Error in one entanglement would not affect the quantum states of other entanglements, resulting in IEDTC performing with an average greater accuracy at network size $3$. Consequently, larger improvements in accuracy are expected with network sizes greater than $3$.

\begin{center}
\begin{figure}[t]
\begin{tikzpicture}
\begin{axis}[
    xlabel={Probability of quantum gate to introduce noise},
    ylabel={Average accuracy},
    xmin=0, xmax=11,
    ymin=.25, ymax=1.0,
    xtick={1, 2, 3, 4, 5, 6, 7, 8, 9, 10},
    legend pos=north east,
    xmajorgrids=true,
    ymajorgrids=true,
]
\addplot[
    color=red,
     error bars/.cd, y dir=both, y explicit,
    ]
    coordinates {
    (1,0.936) +=(0,0.015) -=(0,0.015)
    (2,0.899) +=(0,0.019) -=(0,0.019)
    (3,0.824) +=(0,0.024) -=(0,0.024)
    (4,0.776) +=(0,0.026) -=(0,0.026)
    (5,0.747) +=(0,0.027) -=(0,0.027)
    (6,0.677) +=(0,0.029) -=(0,0.029)
    (7,0.663) +=(0,0.029) -=(0,0.029)
    (8,0.624) +=(0,0.030) -=(0,0.030) 
    (9,0.566) +=(0,0.030) -=(0,0.030) 
    (10,0.533) +=(0,0.031) -=(0,0.031) 
    };
    \addlegendentry{\cite{herbert_ghz}, N=2}
\addplot[
    color=yellow,
         error bars/.cd, y dir=both, y explicit,
    ]
    coordinates {
    (1, 0.883) +=(0,0.02) -=(0,0.02)
    (2,0.745) +=(0,0.027) -=(0,0.027)
    (3,0.680) +=(0,0.029) -=(0,0.029)
    (4,0.613) +=(0,0.03) -=(0,0.03)
    (5,0.526) +=(0,0.031) -=(0,0.031)
    (6,0.460) +=(0,0.031) -=(0,0.031)
    (7,0.407) +=(0,0.03) -=(0,0.03)
    (8,0.383) +=(0,0.03) -=(0,0.03)
    (9,0.328) +=(0,0.029) -=(0,0.029)
    (10,0.313) +=(0,0.029) -=(0,0.029)
    };
    \addlegendentry{\cite{herbert_ghz}, N=3}
\addplot[
    color=blue,
         error bars/.cd, y dir=both, y explicit,
    ]
    coordinates {
    (1,0.925) +=(0,0.016) -=(0,0.016)
    (2,0.896) +=(0,0.019) -=(0,0.019)
    (3,0.839) +=(0,0.023) -=(0,0.023
    (4,0.767) +=(0,0.026) -=(0,0.026)
    (5,0.722) +=(0,0.028) -=(0,0.028)
    (6,0.693) +=(0,0.029) -=(0,0.029)
    (7,0.654) +=(0,0.029) -=(0,0.029)
    (8,0.600) +=(0,0.03) -=(0,0.03)
    (9,0.573) +=(0,0.031) -=(0,0.031)
    (10,0.550) +=(0,0.031) -=(0,0.031)
    };
    \addlegendentry{IEDTC, $N$=2}
\addplot[
    color=green,
         error bars/.cd, y dir=both, y explicit,
    ]
    coordinates {
    (1,0.899) +=(0,0.019) -=(0,0.019)
    (2,0.821) +=(0,0.024) -=(0,0.024)
    (3,0.751) +=(0,0.027) -=(0,0.027)
    (4,0.691) +=(0,0.029) -=(0,0.029) 
    (5,0.634) +=(0,0.03) -=(0,0.03)
    (6,0.578) +=(0,0.031) -=(0,0.031)
    (7,0.497) +=(0,0.031) -=(0,0.031)
    (8,0.465) +=(0,0.031) -=(0,0.031)
    (9,0.447) +=(0,0.031) -=(0,0.031)
    (10,0.355) +=(0,0.03) -=(0,0.03)
    };
    \addlegendentry{IEDTC, $N$=3}
\end{axis}
\end{tikzpicture}
\caption{Accuracy of IEDTC and benchmark \cite{herbert_ghz}.}
\label{fig:accuracy}
\end{figure}
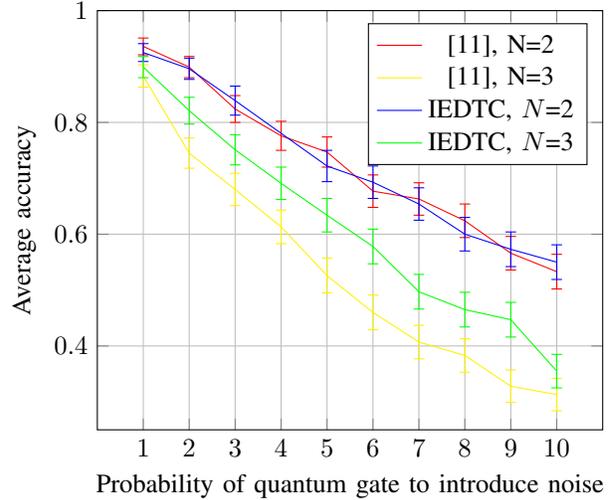
\end{center}

Fig. \ref{fig:timing} shows the time taken in simulation to distribute and teleport quantum states by both IEDTC and the benchmark \cite{herbert_ghz}. Simulation time was taken by averaging the runtime of both protocols over $100$ iterations with $95\%$ confidence interval. IEDTC surpassed the benchmark, achieving an overall $35.2\%$ faster simulation time. The benchmark utilizes a comparable number of qubits at network size $2$, but at this size it requires a greater number of quantum operations, limiting its efficiency. When distributing entanglements, the benchmark transmits $N \times (N-1)$ qubits at startup and cannot continue until all of these qubits reach their destination. As such, the protocol is unable to serve individual entanglements and must distribute them in batches, requiring quantum memory to do so. IEDTC, in comparison, establishes individual entanglements in parallel. $M_2$ and each $T_n$ transmit their required states synchronously to the target receiver, minimizing the need for quantum memory and utilizing parallelism. IEDTC performs with greater efficiency compared to the benchmark when distributing and teleporting states.

\begin{center}
\begin{figure}[t]
\begin{tikzpicture}
\begin{axis}[
    xlabel={Network size},
    ylabel={Time [Seconds]},
    xmin=1.5, xmax=10.5,
    ymin=3, ymax=20,
    xtick={2, 3, 4, 5, 6, 7, 8, 9, 10},
    legend pos=north west,
    xmajorgrids=true,
    ymajorgrids=true,
]

\addplot[
    color=blue,
         error bars/.cd, y dir=both, y explicit,
    ]
    coordinates {
    (2,4.472) +=(0,0.009) -=(0,0.009)
    (3,9.125) +=(0,0.326) -=(0,0.326)	

    };
    \addlegendentry{\cite{herbert_ghz}}

\addplot[
    color=red,
         error bars/.cd, y dir=both, y explicit,
    ]
    coordinates {
    (2,3.638) +=(0,0.009) -=(0,0.009)
    (3,5.458) +=(0,0.024) -=(0,0.024)
    (4,7.451) +=(0,0.069) -=(0,0.069)
    (5,9.135) +=(0,0.067) -=(0,0.067)
    (6,11.196) +=(0,0.091) -=(0,0.091)
    (7,13.389) +=(0,0.072) -=(0,0.072)
    (8,15.576) +=(0,0.122) -=(0,0.122)
    (9,17.307) +=(0,0.161) -=(0,0.161)
    (10,19.455) +=(0,0.158) -=(0,0.158)
    };
    \addlegendentry{IEDTC}
    
\end{axis}
\end{tikzpicture}
\caption{Simulation time of IEDTC and benchmark \cite{herbert_ghz}.}
\label{fig:timing}
\end{figure}
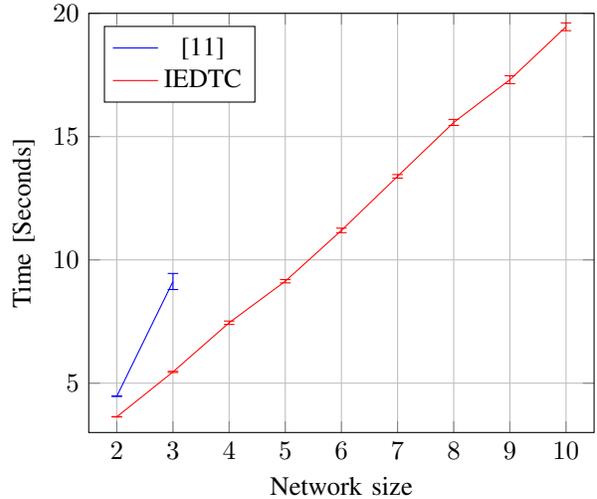
\end{center}

\subsection{QSRE Results}

In Fig. \ref{fig:qsrs-encode-sr}, $4$ bits read (two bits for direction and sign and two bits for the angle $\theta$) from the shared key was sufficient to achieve the lowest probability of $7.2\%$ for guessing how to decode the quantum state. When further bits were introduced we find the probability of guessing the correct counter-rotation increases. From the original equation $\theta$ = \( \displaystyle  \frac{\pi}{-1^{b} \times (1+ d)} \), increasing the number of bits to determine the value of $d$ provides more possible angles that the quantum state can be rotated by. However, the degree by which the quantum state is rotated is decreased as the divisor increases, to the point that certain angles of rotation are close enough that do not affect the state's value with any discernible difference from each other. After a point, increasing the possibility space of $d$ increases the probability space where the malicious receiver is able to guess a value for $\theta$, as more equivalent values can be chosen by the encoded and guessed by the adversary. This results in an increase in false positives, with the malicious receiver acquiring a correct state even when the degree of rotation is guessed incorrectly. This rate increases significantly as further bits are used to determine the angle $\theta$, reducing the effectiveness of this method. As a result, $4$ bits read from the secret key are sufficient to attain the lowest probability of $7.2\%$ that the malicious receiver would correctly recover the teleported state.  


\begin{center}
\begin{figure}[t]
\begin{tikzpicture}
\begin{axis}[
    xlabel={Bits read from private key.},
    ylabel={Percentage of states retrieved},
    xmin=2, xmax=9,
    legend pos=north west,
    xmajorgrids=true,
    ymajorgrids=true,
]

\addplot[
    color=blue,
             error bars/.cd, y dir=both, y explicit,
    ]
    coordinates {

    (3,0.092)  +=(0,0.0252) -=(0,0.0252)
    (4,0.072)  +=(0,0.0226) -=(0,0.0226)
    (5,0.076)  +=(0,0.0232) -=(0,0.0232)
    (6,0.086)  +=(0,0.0246) -=(0,0.0246)
    (7,0.156)  +=(0,0.0318) -=(0,0.0318)
    (8,0.336)  +=(0,0.0414) -=(0,0.0414)

    };
    \addlegendentry{Success rate}
    
\end{axis}
\end{tikzpicture}
\caption{Success rate of eavesdropper decoding quantum state.}
\label{fig:qsrs-encode-sr}
\end{figure}
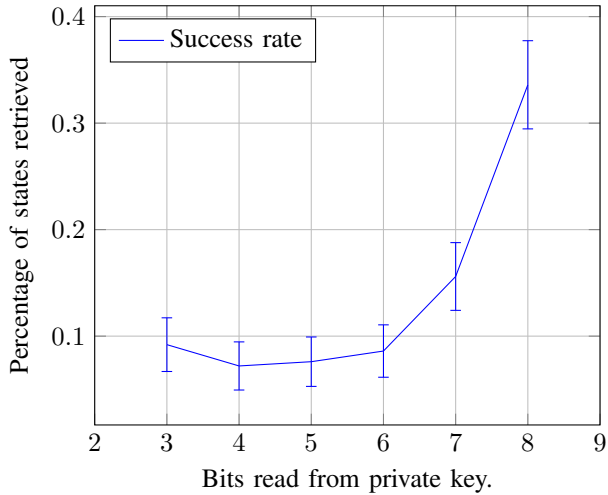
\end{center}

\section{Conclusions}
\label{sec:conclusions}

This paper proposes IEDTC, an entanglement distribution and quantum teleportation protocol that can distribute entangled pairs between indirectly connected nodes and avoid bottlenecks that would reduce the teleportation throughput. To accomplish this, IEDTC distributes entanglements by entanglement swapping, and then utilizes quantum teleportation with classical network coding. By using classical links and a limited number of quantum links to exchange quantum states, we enable near-term classical-quantum networks for distributed quantum computing. To gauge our performance, IEDTC and a state-of-the-art benchmark were implemented in QuNetSim. Simulation results demonstrate that IEDTC requires fewer qubits and network links, achieves greater efficiency and accuracy, and scales better than the benchmark. 
To support the security of IEDTC against malicious entanglement attacks, we propose QSRE. As encrypting a $2$-bit teleportation message has a maximum gain of $1/4$ to guess the original message, we instead proposed encoding the teleported quantum state through angular rotation. QSRE successfully lowered the success rate of a malicious receiver eavesdropping on this state without incurring additional overhead, the only requirement being a pre-shared private key for each transceiver pair. As the key's contents are not leaked through the encoding process, transceiver pairs can continue using the same key while communicating with each other.

\ifCLASSOPTIONcaptionsoff
  \newpage
\fi

\bibliographystyle{IEEEtran} 
\bibliography{refs} 

\end{document}